\documentclass[aps,prl,twocolumn,showpacs,superscriptaddress]{revtex4-1}
\usepackage{bm}
\usepackage{mathrsfs}
\usepackage{amsmath}
\usepackage{amssymb}
\usepackage{graphicx}
\usepackage{amsfonts}
\usepackage{amsthm}
\usepackage{color}
\usepackage{dcolumn}
\usepackage{txfonts}

\begin{document}

\title{Probing Yang-Lee Edge Singularity by Central Spin Decoherence}
\author{Bo-Bo Wei}
\email{Corresponding author: bbwei@szu.edu.cn}
\affiliation{School of Physics and Energy, Shenzhen University, Shenzhen 518060, China}

\begin{abstract}
Yang-Lee edge singularities are the branch point of the free energy on the complex plane of physical parameters and were shown to be the simplest universality class of phase transitions. However, the Yang-Lee edge singularities have not been regarded as experimentally observable since they occur at complex physical parameters, which are unphysical. A recent discovery about the relation between partition functions and probe spin coherence makes it experimentally feasible to access the complex plane of physical parameters. However, how to extract the critical point and the critical exponent of Yang-Lee edge singularities in many-body systems, which occurs only at thermodynamic limit, has still been elusive.  Here we show that the quantum coherence of a probe spin coupled to finite-size Ising-type spin systems presents universal scaling behavior near the Yang-Lee edge singularity. The finite-size scaling behavior of quantum coherence of the probe spin predicts that one can extract the critical point and the critical exponent of the Yang-Lee edge singularity of Ising-type spin system in the thermodynamic limit from the spin coherence measurement of the probe spin coupled to finite Ising-type spin systems. This finding provides a practical approach to studying the nature of Yang-Lee edge singularities of many-body systems.
\end{abstract}


\maketitle

In 1952, C. N. Yang and T. D. Lee \cite{LY1952a,LY1952b} initiated to examine phase transitions by studying the partition function zeros, termed as Lee-Yang zeros, on the complex plane of magnetic field $h$ (or fugacity $z$ for fluids). They found that for a system above its critical temperature $T_c$, the partition function must be nonzero throughout some neighborhood of the real axis of magnetic field within a gap determined by two edges at $\pm h_{YL}(T)$, therefore the free energy is an analytic function of the real magnetic field. On the other hand, below the critical temperature $T_c$, the Lee-Yang zeros will come arbitrarily to the real axis as the thermodynamic limit is taken, destroying the analyticity of the free energy in $h$ for those real fields at which the Lee-Yang zeros accumulate.

Kortman and Griffiths \cite{Kortman1971} pointed out that the edges of Lee-Yang zeros of a many-body system in the thermodynamic limit are singularity points, termed Yang-Lee edge singularities \cite{Fisher1978}. Fisher \cite{Fisher1978} proposed that the Yang-Lee edge singularity could be considered as a new second order phase transition point with associated critical exponents. Renormalization group analyses \cite{Fisher1978} show that the Yang-Lee edge singularity is associated to a $i\varphi^3$ theory and the crossover dimensionality of the Yang-Lee edge singularity is $d_c=6$. Later, the investigation of the Yang-Lee edge singularity has been extended to many other models such as the classical $n$-vector model \cite{Kurtze1979}, the quantum Heisenberg model \cite{Kurtze1979}, the spherical model \cite{Kurtze1978}, the quantum one-dimensional transverse Ising model \cite{YL1}, the hierarchical model \cite{YL2}, branched polymers \cite{YL3}, the directed-site animals-enumeration problem \cite{YL4}, Ising models on fractal lattices \cite{YL5}, Ising systems with correlated disorder \cite{YL6}, fluid models with repulsive-core interactions \cite{Lai1995}, the antiferromagnetic Ising model \cite{Kim2005} etc. A comprehensive review on the study of Yang-Lee edge singularity can be found at \cite{Review2005,Review2015}. Recently Kibble-Zurek scaling was shown to appear at Yang-Lee edge singularity \cite{Yin2016}, which supports that the Yang-Lee edge singularity is similar to conventional second order phase transitions.

Yang-Lee edge singularity occurs in the ferromagnetic Ising models above its critical temperatures in a purely imaginary magnetic field $ih''$ (Fig.1). For $h''>h_{YL}(T)$, the partition function acquires zeros, which becomes dense on the line with $\Im h>h_{YL}(T)$. While the density of Lee-Yang zeros $g(T,h'')$ in the thermodynamic limit behaves as $g(T,h'')\sim |h''-h_{YL}(T)|^{\sigma}$  when $h''\rightarrow h_{YL}(T)$ from above with $\sigma$ being the universal critical exponent of the Yang-Lee edge singularity. Fisher showed that \cite{Fisher1980} for one-dimensional systems with classical finite range interactions, the critical exponent of Yang-Lee edge singularity takes a universal value $\sigma=-1/2$, independent of the temperature and the details of the Hamiltonian. Cardy showed that \cite{Cardy1986} $\sigma=-1/6$  for Yang-Lee edge singularity in two-dimensional systems. The exponent $\sigma=1/2$ for dimensionality $d\geq d_c$ \cite{Fisher1978}. From the density of Lee-Yang zeros, we know that the free energy density near the Yang-Lee edge \cite{Review2005,Review2015} behaves as $f(T,h'')\sim A_{\pm}|h''-h_{YL}(T)|^{\sigma+1}$  where the amplitudes $A_{\pm}$ may depend on the sign of $h''-h_{YL}(T)$. It follows that the susceptibility, $\chi\sim \partial^2 f/\partial h^2$  diverges with an exponent $\gamma=1-\sigma$.  The correlation function of two spin with distance $R$ apart near the Yang-Lee edge singularity for an Ising type system behaves as  $G(R,T,h'')\sim D(R\tilde{h}^{\nu})/R^{d-2+\eta}$ when $\tilde{h}\equiv h''-h_{YL}(T)\rightarrow0$ and $R\rightarrow\infty$. Similarly, hyperscaling relations, $\sigma=(d-2+\eta)/(d+2-\eta)$  and  $\gamma=2/(d+2-\eta)$, are verified up to upper critical dimension $d_c=6$ \cite{Fisher1978,Kurtze1979,Kurtze1978}.

Although Yang-Lee edge singularity appears ubiquitous in many-body systems, experimental observation of Yang-Lee edge singularities, however, has not been made before. The previous experiments could only indirectly derive the densities of Lee-Yang zeros from susceptibility measurement plus analytic continuation \cite{Binek1998,Binek2001}. The difficulty is intrinsic: The Yang-Lee edge singularities would occur only at complex values of external fields, which are unphysical. A recent theoretical discovery \cite{Wei2012} about the relation between partition functions and probe spin coherence makes it experimentally feasible to access the complex plane of physical parameters and more generally the thermodynamic on the complex plane \cite{Wei2014,Wei2015}. Wei and Liu \cite{Wei2012} found that the quantum coherence of a central spin embedded in an Ising-type spin bath is equivalent to the partition function of the Ising spin bath under a complex magnetic field. Then the Lee-Yang zeros of the partition function are one-to-one mapped to the zeros of the central spin coherence, which are directly measurable. This leads to the first experimental observation of Lee-Yang zeros \cite{Peng2015,LYExp2015}. However how to locate the critical points and the critical exponents of the Yang-Lee edge singularity, which occurs in the thermodynamic limit of a many-body system, has still been elusive.

In this Letter we show that the quantum coherence of the probe spin coupled to finite-size many-body system presents universal scaling behavior around the Yang-Lee edge singularity. The universal finite-size scaling of the quantum coherence of the probe spin predicts the critical point and the critical exponent of the Yang-Lee edge singularity accurately and systematically. By measuring quantum coherence of the probe spin which is coupled to finite-size systems, one can extract the critical point and the critical exponent of Yang-Lee edge singularity of infinite system and verify the universality of the Yang-Lee edge singularity.

Let us consider a general Ising model with ferromagnetic interactions $J_{ij}\geq0$ under
a magnetic field $h$. The Hamiltonian is
\begin{equation}\label{ham}
H(h)=-\sum_{i,j}J_{ij}s_is_j-h\sum_js_j\equiv H_0+hH_1,
\end{equation}
where the spins $s_j$ take values $\pm1$.  We use a probe spin-1/2 coupled to the Ising system (bath), with probe-bath interaction
$H_I=-\lambda S_z\otimes\sum_js_j\equiv\lambda S_z\otimes H_1 $ and $\lambda$ being a coupling constant and $S_z\equiv(|\uparrow\rangle\langle\uparrow|-|\downarrow\rangle\langle\downarrow|)/2$
being the Pauli matrix of the probe spin. If we initiate the probe spin in a superposition state as
$(|\uparrow\rangle+|\downarrow\rangle)/\sqrt{2}$ and the bath at inverse temperature
$\beta=1/T$ descried by $\rho_0=e^{-\beta H}/Z(\beta,h)$ with $Z(\beta,h)=\text{Tr}[e^{-\beta H}]$ being the partition function.  Then the quatum coherence of the probe spin, defined as $L(t)=\langle S_x\rangle+i\langle S_y\rangle$, has an intriguing form as \cite{Wei2012,Wei2014}
\begin{eqnarray}\label{central1}
L(t)=\frac{Z(\beta,h-it\lambda/\beta)}{Z(\beta,h)}.
\end{eqnarray}
The denominator in the above equation is nonzero for real magnetic
field and temperature. The numerator resembles the form of a partition
function but with a complex magnetic field $h-it\lambda/\beta$. The
probe spin coherence in a finite system vanishes whenever $h-it\lambda/\beta$ reaches a Lee-Yang zero. Particularly for Ising
ferromagnets, the Lee-Yang zeros all lie on the unit circle, the $n$-th Lee-Yang zeros $h_n=ih''_n,n=1,2,\cdots,N$, and
therefore are mapped to the probe spin coherence zeros ($t_n$) for
vanishing external field ($h=0$), with the correspondence relation $t_n=h''_n/\lambda$. This leads to the first experimental observation of Lee-Yang zeros \cite{Peng2015,LYExp2015}.

\begin{figure}
\begin{center}
\includegraphics[width=\columnwidth]{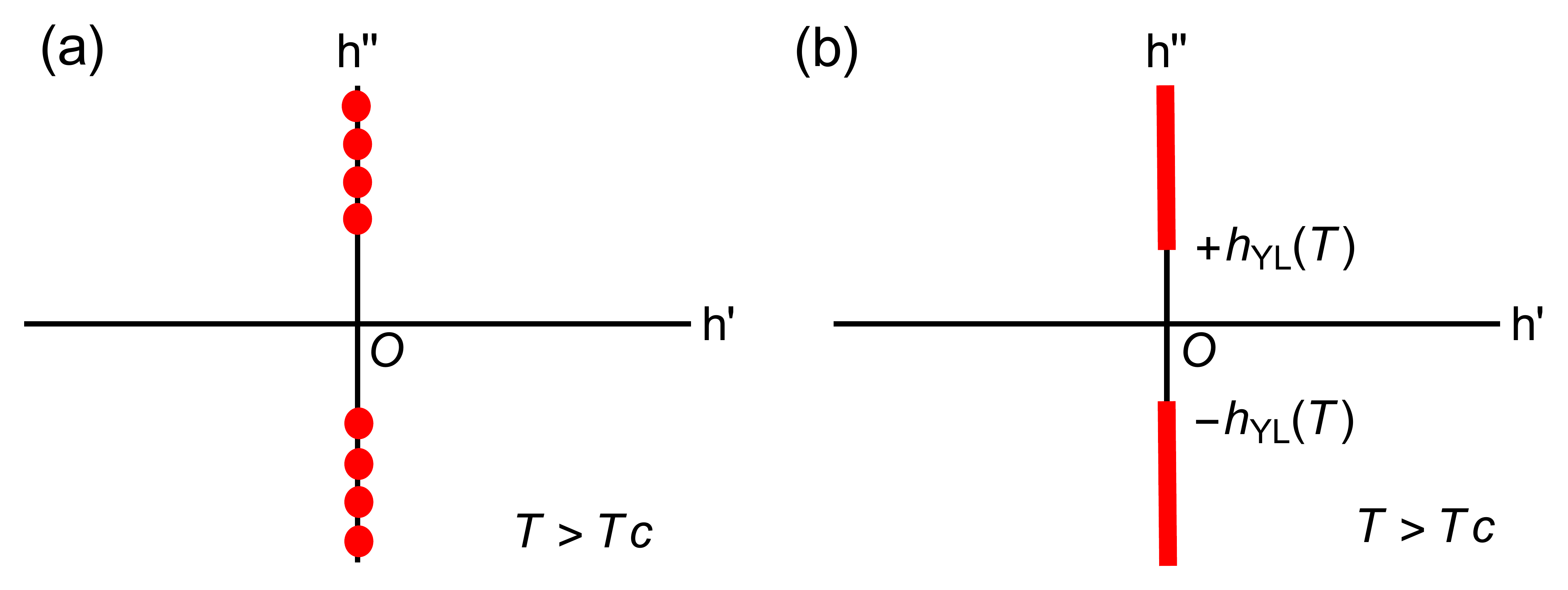}
\end{center}
\caption{(color online). Schematic illustration of Lee-Yang zeros and Yang-Lee edge singularities in a general Ferromagnetic Ising models on the complex plane of magnetic field $h=h'+ih''$.  (a). The Lee-Yang zeros in a finite-size Ferromagnetic Ising model where unit circle theorem holds, i.e. all the zeros are purely imaginary.  (b). As the thermodynamic limit is taken, the discrete Lee-Yang zeros condensed into a line with the two edge termed Yang-Lee edge at $\pm h_{YL}(T)$  when temperature is above the critical temperature.}
\label{fig:epsart1}
\end{figure}

To probe Yang-Lee edge singularities, which occurs only at thermodynamic limit, is non-trivial because any realistic experiment is always performed on finite-sized samples. To extract the properties of an infinite system from the experimental data for finite-size systems, one needs to employ finite-size scaling technique invented by Fisher and Barber \cite{Fisher1972}. The central idea of finite-size scaling is that the physical observable of a finite-sized system with spatial extent characterized by a length $l$ only depends on the universal ratio $l/\xi$ with $\xi$ being the correlation legnth of the infinite system and it has been an important tool for the understanding and development of statistical mechanics of systems which are close to a critical point \cite{Cardy1988}.

Applying the Finite-size scaling hypothesis \cite{Fisher1972} to Yang-Lee edge singularities, we have the free energy per spin of a finite system $f_l$ with spatial extension $l$ present finite-size scaling behaviour,
\begin{eqnarray}
\frac{f_l(\tilde{h})}{f_{\infty}(\tilde{h})}&=&\Phi(l/\xi)
\end{eqnarray}
where $\tilde{h}=h'+ih''-h_c(T)$ and $h_c(T)$ being the position of the Yang-Lee edge singularity and $\xi\sim \tilde{h}^{-\nu}$ is the correlation length near the Yang-Lee edge singularitity for infinite system and $\Phi(x)$ is a universal function. Because the free energy density near the Yang-Lee edge singularitity for infinite system behaves as $f_{\infty}(\tilde{h})\sim \tilde{h}^{\sigma+1}$ \cite{Review2005,Review2015} and the hyperscaling relation $\nu=(\sigma+1)/d$, we have the free energy density of a finite system $f_l$ present finite-size scaling behaviour \cite{Review2005,FSS1},
\begin{eqnarray}
l^df_l(\tilde{h})=\tilde{\Phi}(\tilde{h}l^{d/(\sigma+1)}),
\end{eqnarray}
where $d$ is the dimensionality of the system and $\sigma$ is the univeral critical exponent associated to the Yang-Lee edge singularity and $\tilde{\Phi}(x)$ is a new universal function. The relation between probe spin coherence and partition function with a complex parameter in Eq.\eqref{central1} implies that the rescaled probe spin coherence coupled to an Ising type spin bath present the finite-size scaling behavior,
\begin{eqnarray}\label{central0}
\tilde{L}(t)&=&L(t)Z(\beta,h)=\tilde{\Psi}\Big((h-i\lambda t/\beta-h_c)l^{d/(\sigma+1)}\Big).
\end{eqnarray}
Here $\tilde{\Psi}(x)$ is a universal scaling function. Particularly, for Ising ferromagnets, the Yang-Lee edge singularity is purely imaginarty $h_c=ih_{YL}(T)$ and therefore
we study the probe spin coherence for the vanishing external field $h=0$. According to Finite-size scaling hypotheis, the rescaled probe spin coherence which is coupled to an Ferromagnetic Ising type spin bath at zero magnetic field presents a universal finite-scaling behaviour,
\begin{eqnarray}\label{central2}
\tilde{L}(t)&=&L(t)Z(\beta,0)=\Psi\Big(\lambda(t-t_c)l^{d/(\sigma+1)}\Big).
\end{eqnarray}
Here $\Psi(x)$ is a universal scaling function and analytic for finite size systems and $t_c=\beta h_{YL}/\lambda$. This universal finite-size scaling behaviour of quantum coherence is powerful for studying the nature of Yang-Lee edge singularities: \\
(1).~\emph{Identifying the critical point of Yang-Lee edge singularity}: If we plot $\tilde{L}(t)$  as a function of time for systems with different sizes at a fixed temperature $T$, all the curves cross at $t=t_c=\beta h_{YL}(T)/\lambda$ and $\tilde{L}(t_c)=\Psi(0)$. From the crossing point of $\tilde{L}(t)$  for different system sizes with a fixed temperature, we therefore can locate the Yang-Lee edge singularity for infinite system. \\
(2).~\emph{Data collapse}: If we draw $\tilde{L}(t)$ as a function of $\lambda(t-t_c)l^{d/(\sigma+1)}$, then the curves for different system sizes collapse around the Yang-Lee edge singularity because $\Psi(x)$ is a universal scaling function. \\
(3).~\emph{Extracting the universal critical exponent of Yang-Lee edge singularity}: Since the universal scaling function $\Psi(x)$ is an analytic function for finite-size systems, Taylor expansion of $\Psi(x)$ around $x=0$ in Eq.\eqref{central2} leads to
\begin{eqnarray}\label{central3}
\tilde{L}(t)&=&\Psi(0)+\Psi'(0)\lambda(t-t_c)l^{d/(\sigma+1)}+o(t-t_c)^2.
\end{eqnarray}
Differentiating both sides of Eq. \eqref{central3} with respect to $\lambda t$ and taking $t=t_c$, we obtain
\begin{eqnarray}\label{central4}
\tilde{L}(t_c)'&=&\Psi'(0) l^{d/(\sigma+1)}.
\end{eqnarray}
Here $\tilde{L}(t_c)'\equiv\frac{1}{\lambda}\frac{d\tilde{L}(t)}{d t}|_{t=t_c}$. Taking logarithm on both sides of Eq. \eqref{central4}, we get
\begin{eqnarray}\label{central5}
\ln\tilde{L}(t_c)'&=&\ln\Big(\Psi'(0)\Big)+\frac{d}{\sigma+1}\ln l.
\end{eqnarray}
Eq.\eqref{central5} predicts that $\ln\tilde{L}(t_c)'$ is a linear function of $\ln l$ with slop being $d/(\sigma+1)$. Thus we can also extract the critical exponent for the Yang-Lee edge singularities from the probe spin coherence measurement for finite-size systems.

Note that in Eq.~\eqref{central2}, it is not the central spin coherence $L(t)$ but the rescaled central spin coherence $\tilde{L}(t)$, which is equal to the central spin coherence times partition function at zero magnetic field presents finite-size scaling behavior. Since the Yang-Lee edge singularity only occurs for temperature above the critical temperature, then one can obtain the partition function at zero magnetic field from high temperature expansion method \cite{Stanley1971,Wipf2013}. An alternative method to get the partition function of a finite system at high temperatute is the cluster correlation expansion method \cite{Yang2008,Yang2009}.

\begin{figure}
\begin{center}
\includegraphics[width=\columnwidth]{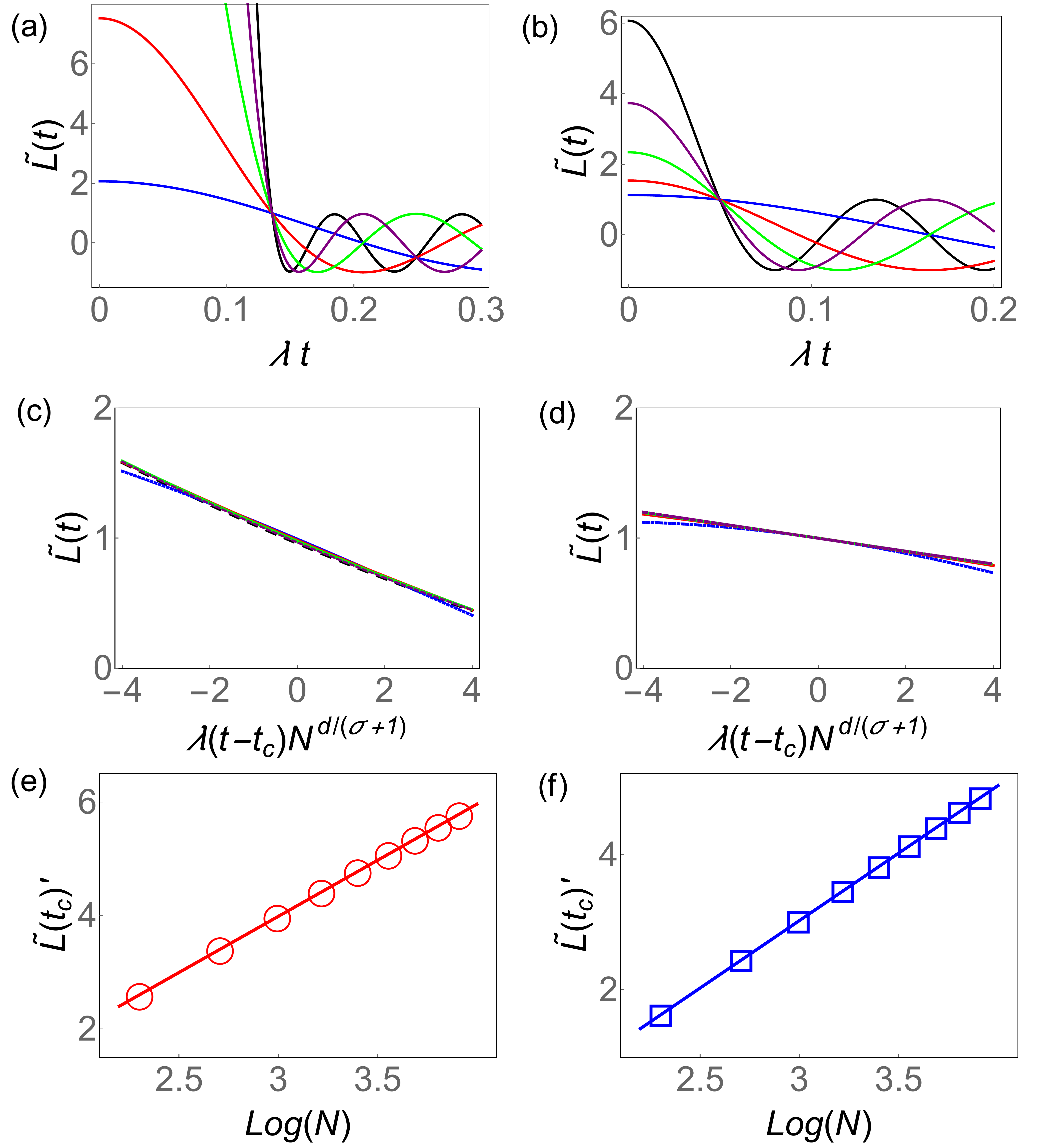}
\end{center}
\caption{(color online). Yang-Lee edge singularity in 1D Ising model. (a). The rescaled spin coherence $\tilde{L}(t)$ as a function of time in 1D Ising model at inverse temperature $\beta=1.0$ for systems with different number of spins $N=10,20,30,40,50$ from bottom to top, respectively. (b). The same as (a) but for inverse temperature $\beta=1.5$. (c). The collapse of the curves in (a) when $t$ is rescaled by the critical exponents of Yang-Lee edge singularity in 1D Ising model, $¦Ë(t-tc)N^{d/(\sigma+1)}$. (d). The same as (c) but for inverse temperature $\beta=1.5$. (e). The logarithm of the first derivative of the scaled spin coherence at critical point $t_c$ as a function of the logarithm of the system sizes $\ln N$. (f). The same as (e) but for inverse temperature $\beta=1.5$.}
\label{fig:epsart2}
\end{figure}

\begin{figure}
\begin{center}
\includegraphics[width=\columnwidth]{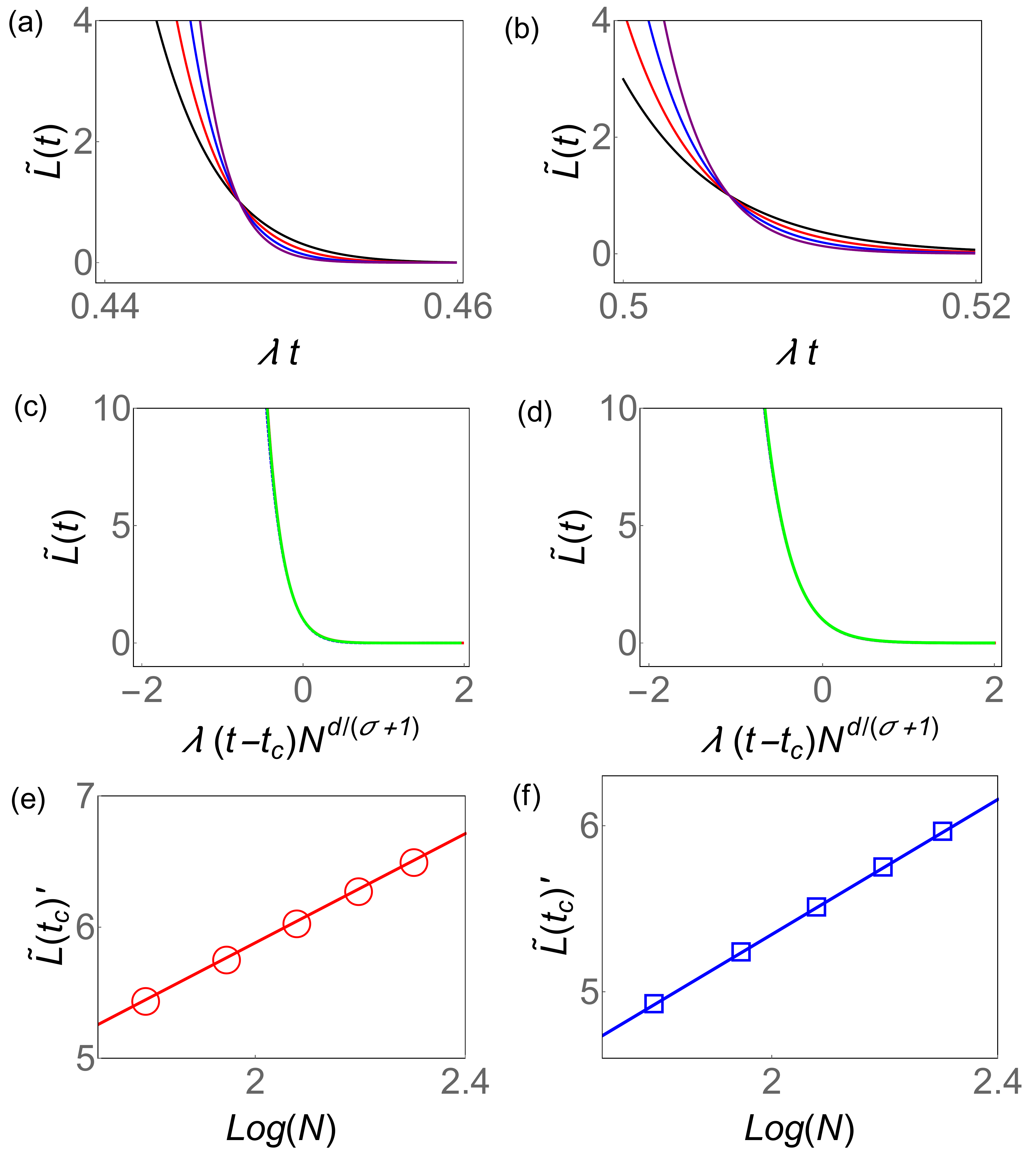}
\end{center}
\caption{(color online). Yang-Lee edge singularity in 2D Ising model. (a). The rescaled spin coherence $\tilde{L}(t)$ as a function of time in 2D Ising model at inverse temperature $\beta=0.04$ for systems with different number of spins $N\times N=7\times 7,8\times 8,9\times 9,10\times 10$ from bottom to top, respectively. (b). The same as (a) but for inverse temperature $\beta=0.01$. (c). The collapse of the curves in (a) when $t$ is rescaled by the critical exponents of Yang-Lee edge singularity in 2D Ising model, $¦Ë(t-tc)N^{d/(\sigma+1)}$. (d). The same as (c) but for inverse temperature $\beta=0.01$. (e). The logarithm of the first derivative of the scaled spin coherence at critical point $t_c$ as a function of the logarithm of the system sizes $\ln N$ at $\beta=0.04$. (f). The same as (e) but for inverse temperature $\beta=0.01$.}
\label{fig:epsart3}
\end{figure}

To illustrate the above idea, we study the one-dimensional (1D) Ising model with nearest-neighbor ferromagnetic coupling $J=1$  and the periodic boundary condition. The 1D Ising model can be exactly solved through the transfer matrix method \cite{Kramers1941} and there is no finite temperature phase transition in 1D Ising model, i.e. $T_c=0$.  Figure 2 shows the extracting the Yang-Lee edge singularity in 1D Ising model from probe spin coherence measurement. At inverse temperature $\beta=1.0$ , which is above the critical temperature of 1D Ising model, the rescaled probe spin coherence for systems with different sizes $N=10, 20, 30, 40, 50$, cross at point $t=t_c$ as predicted from finite-size scaling theory [Fig.~1(a)]. When we make a scale transformation of the probe spin coherence by the universal critical exponent of Yang-Lee edge singularity in 1D system, the curves in Fig.~1(a) for different system sizes collapse around the Yang-Lee edge singularity [Fig.~1(c)]. In Fig.~1 (e), we plot $\ln(\tilde{L}(t_c)')$ as a function of $\ln N$  at inverse temperature $\beta=1.0$. According to the prediction of finite-size scaling theory Eq.\eqref{central5}, they are linearly related with the slope being $d/(\sigma+1)$ in the 1D Ising model. A linear extropolation method shows that $\sigma=-0.495$.  We can see the finite-size predictions matches the exact solution $\sigma=-1/2$ very well. With temperature decreasing but still above the critical temperature, the Yang-Lee edge singularity point moves towards the real axis. As shown in Fig.~1(b), the rescaled probe spin coherence for different system sizes $N=10, 20, 30, 40, 50$ at inverse temperature $\beta=1.5$ cross at $t=t_c$. If we rescaled the data by the universal critical exponent of Yang-Lee edge singularity in 1D, the curves in Fig.~1(b) collapse [Fig.~1(d)]. In Fig.~1(f), we show $\ln(\tilde{L}(t_c)')$ as a function of $\ln N$ at inverse temperature $\beta=1.5$. One can see that they are linearly related and an extropolation method shows that $\sigma=-0.498$.  We can see the finite-size predictions matches the exact solution $\sigma=-1/2$ very well again. Thus it is feasible to extract the critical point and the universal critical exponent of Yang-Lee edge singularities from central spin coherence measurement.

We further study the two-dimensinal (2D) Ising model with nearest-neighbor ferromagnetic coupling $J=1$ and the periodic boundary condition.  The 2D Ising model has been exactly solved by Onsager in 1944 \cite{Onsager1944} and there is a finite temperature phase transition at $\beta_c\approx 0.44$ \cite{Onsager1944}. For 2D Ising model under a finite magnetic field, there is no exact solution available but one can map the problem into 1D quantum Ising model with both longitudinal and transverse field by transfer matrix method \cite{Schultz1964}. Figure 3 shows the extracting the Yang-Lee edge singularity in the 2D Ising model from central spin coherence measurement. At inverse temperature $\beta=0.04$, which is above the critical temperature of the 2D Ising model, the rescaled probe spin coherence data for different systems sizes $N\times N=7\times 7, 8\times8, 9\times9, 10\times10$, respectively, cross at point $t=t_c$ [Fig.~3(a)]. In Fig.~3(e), we present $\ln(\tilde{L}(t_c)')$ as a function of $\ln N$ at inverse temperature $\beta=0.04$. An extropolation of the data shows that $\sigma=-0.05$, which is a little bit further away from the exact solution $\sigma=-1/6$. This is because the system size we studied is too small. We then plot $\tilde{L}$  as a function of  $\lambda(t-t_c)N^{d/(\sigma+1)}$ with $d=2,\sigma=-0.05$ for different system sizes and it turns out that all the curves collapse perfectly [Fig.~3(c)].  The Yang-Lee edge singularity point moves towards the real axis as temperature decreases. We show $\tilde{L}(t)$ as a function of time for different system sizes, $N\times N=7\times 7, 8\times8, 9\times9, 10\times10$, respectively at inverse temperature $\beta=0.01$ in Fig.~3(b). One can see that the rescaled probe spin coherence for different systems sizes cross at point $t=t_c$, which is the Yang-Lee edge singularity of the 2D Ising model. In Fig. 3(f), $\ln(\tilde{L}(t_c)')$ as a function of $\ln N$ at inverse temperature $\beta=0.01$. We find that the critical exponenet of the Yang-Lee edge singularity in 2D Ising model is $\sigma=-0.052$ by extropolation method. We then plot $\tilde{L}$  as a function of  $\lambda(t-t_c)N^{d/(\sigma+1)}$ with $d=2,\sigma=-0.052$ for different system sizes and it turns out that all the curves collapse completely [Fig.~3(d)]. These results prove the feasibility of the central spin decoherence measurement for studying the nature of Yang-Lee edge singularity.

In summary, we show that the quantum coherence of a probe spin coupled to a finite-size Ising-type spin bath presents universal finite-size scalng behavior near the Yang-Lee edge singularity. The universal finite-size scaling of the quantum coherence of the probe spin predicts the critical point and the critical exponent of the Yang-Lee edge singularity. Thus measuring quantum coherence of a probe spin which is coupled to finite-size many-body systems, one can extract the critical point and the critical exponent of Yang-Lee edge singularity of the infinite system. In particular for ferromagnetic Ising models, the unit-circle theorem holds regardless of the interaction range, geometry configurations, disorders, and dimensionality. Such universality offers a great deal of feasibility and flexibility for experimental observation of the Yang-Lee edge singularities. For other systems (e.g., anti-ferromagnetic Ising models), the Yang-Lee edge singularities may not lie on the imaginary axis. But one can apply an external field $h$, and get the Yang-Lee edge singularity of with real part $h$. Thus measuring quantum coherence of a single spin provides a universal tool to probe the singularity point of many-body systems on the complex plane of physical parameters and more general the thermodynamics on the complex plane. In the complex plane of temperature, there are also singularities \cite{Fisher1965}, termed as Fisher edge singularities, thus one may expect similar finite-size scaling techniques can be applied to probe the universal nature of the Fisher edge singularities.

\begin{acknowledgements}
This work was supported by the National Natural Science Foundation of China (Grant no. 11604220) and the Startup Fund of Shenzhen University (Grant no. 2016018).
\end{acknowledgements}

\end{document}